# Octave-spanning supercontinuum generation in a CMOS-compatible thin Si$_3$N$_4$ waveguide coated with highly nonlinear TeO$_2$

Hamidu M. Mbonde,[1] Neetesh Singh,[2] Bruno L. Segat Frare,[1] Milan Sinobad,[2] Pooya Torab Ahmadi,[1] Batoul Hashemi,[1] Dawson. B. Bonneville,[1] Peter Mascher,[1] Franz X. Kärtner,[2,3] and Jonathan D. B. Bradley[1]

[1]*Department of Engineering Physics, McMaster University, 1280 Main Street West, Hamilton, ON L8S 4L7, Canada*
[2]*Centre for Free Electron Laser Science (CFEL)-DESY and University of Hamburg, Notkestrasse 85, 22607 Hamburg, Germany*
[3]*Department of Physics, Universität Hamburg, Jungiusstraße 9, 20355 Hamburg, Germany*
*\*Corresponding author: mbondeh@mcmaster.ca*

**Supercontinuum generation (SCG) is an important nonlinear optical process enabling broadband light sources for many applications, for which silicon nitride (Si$_3$N$_4$) has emerged as a leading on-chip platform. To achieve suitable group velocity dispersion and high confinement for broadband SCG the Si$_3$N$_4$ waveguide layer used is typically thick (>~700 nm), which can lead to high stress and cracks unless specialized processing steps are used. Here, we report on efficient octave-spanning SCG in a thinner moderate-confinement 400-nm Si$_3$N$_4$ platform using a highly nonlinear tellurium oxide (TeO$_2$) coating. An octave-spanning supercontinuum is achieved at a low peak power of 258 W using a 100-fs laser centered at 1565 nm. Our numerical simulations agree well with the experimental results showing an increase of waveguide's nonlinear parameter by 2.5× when coating the Si$_3$N$_4$ waveguide with TeO$_2$ film. This work demonstrates highly efficient SCG via effective dispersion engineering and an enhanced nonlinearity in a CMOS-compatible hybrid TeO$_2$-Si$_3$N$_4$ waveguides and a promising route to monolithically integrated nonlinear, linear, and active functionalities on a single silicon photonic chip. © 2023 Optica Publishing Group**

Supercontinuum generation (SCG) involves the spectral broadening of short and narrow-band optical pulses that can expand to cover an octave or multiple octaves, leading to broadband light sources for a tremendous variety of applications in communications, frequency metrology, spectroscopy, and imaging [1]. Initially, SCG was extensively studied in optical fibers [2], but recent research has intensively focused on generating supercontinuum (SC) in integrated photonic chips that can offer high performance at more compact form factors.

Significant efforts have been directed towards silicon-based complementary metal oxide semiconductor (CMOS) compatible platforms, which promise cost-effective and compact devices that can be mass-produced and co-integrated with electronic circuits [1]. It is also imperative to generate over an octave of SC and at near-infrared (NIR) wavelengths. Octave spanning SC spectra are critical for self-stabilization of optical frequency combs using *f*-2*f* interferometry enabling precise measurement of absolute optical frequencies [3]. Operation in the NIR allows the use of readily available laser sources and applications in telecommunication wavelengths.

On-chip SCG covering at least an octave at NIR has been demonstrated in various CMOS platforms such as silica [4], silicon-on-insulator (SOI) [5–7], tantalum pentoxide (Ta$_2$O$_5$) [8], silicon-rich nitride (SRN) [9], and silicon nitride (Si$_3$N$_4$). Each platform has tradeoffs in terms of nonlinearity, optical confinement, linear and nonlinear losses, and processing capabilities. With its properties of CMOS compatibility, ultra-low losses, negligible two-photon absorption (TPA), and fabrication maturity and versatility, Si$_3$N$_4$ is the most attractive platform. In the last decade, over-octave and coherent SCs have been demonstrated by various research groups [10–19]. Despite extensive demonstrations of SCG in Si$_3$N$_4$, to achieve an octave spanning bandwidth the required power and energy has been high with high repetition rate sources [20], a constraint imposed by its relatively lower nonlinear index ($n_2$) of about 2.4 × 10$^{-19}$ m$^2$/W at 1550 nm [21]. Another constraint arises from the difficulties of fabricating low-loss thick stoichiometric Si$_3$N$_4$ waveguides, which are required for anomalous dispersion to enable phase-matching and efficient SCG, using conventional thin film deposition and etching methods. This is due to cracks caused by tensile stress in the film introduced during the low-pressure chemical vapor (LPCVD) deposition. Thus, it has necessitated the development of custom foundry processes to achieve exceptional performance such as the photonic damascene [22], stress release patterning [23], thermal cycling [24], and diced trenches techniques [25].

Tellurium dioxide (TeO$_2$) is an oxide glass with very attractive linear and nonlinear optical properties. It has been widely studied in fibers for rare-earth doped amplifiers and lasers, broadband Raman amplification, and SCG [26]. Recently, TeO$_2$ has also been explored for integrated planar waveguide applications. TeO$_2$ glass has a wide transparency spanning from visible to mid-infrared wavelengths and a high refractive index allowing the fabrication of compact waveguides. It is a highly nonlinear medium with $n_2$ of up to 20× that of silica, the highest reported Raman gain coefficient among all oxide glasses, and a large acousto-optic coefficient [27]. Furthermore, TeO$_2$ has proven to be an excellent host of rare earth ions where integrated rare earth doped amplifiers and lasers have been demonstrated in a standalone TeO$_2$ [28,29], and hybrid TeO$_2$-coated Si$_3$N$_4$ and SOI waveguides [30–32]. SCG in a TeO$_2$ waveguide was first studied by S. J. Madden and K. T. Vu [27] showing ~50 nm spectral broadening in a 4-μm-wide and 0.8-μm-high rib waveguide etched on a 1.8-μm thick TeO$_2$ film. Recently, we demonstrated a SC spectrum extending over 500 nm, covering the entire telecom band [33]. The SC was generated in a normal dispersion regime with a hybrid waveguide consisting of a 200-nm thick Si$_3$N$_4$ strip coated with a 370-nm-thick TeO$_2$ layer.

In this work, we present the observation of an octave-spanning SCG in a wafer-scale, CMOS-compatible, crack-free 400-nm-thick Si$_3$N$_4$ waveguide platform coated with highly nonlinear TeO$_2$. The observed SC reaches an octave for a pulse peak power of just 258 W which to the best of our knowledge is the lowest power for an octave SC in a commercial Si$_3$N$_4$ platform pumped at telecom wavelengths. The ~400 nm Si$_3$N$_4$ is a standard thickness fabricated in wafer-scale platforms offered by commercial silicon nitride [34] and silicon photonics [35] foundries. The added TeO$_2$ serves the role of engineering the dispersion of the thin Si$_3$N$_4$ into the anomalous regime while also enhancing the nonlinearity owing to its higher $n_2$ [36]. This platform has the potential of providing an easy route to compact photonic circuits consisting of nonlinear, passive, and active functionalities in one chip.

The Si$_3$N$_4$ chip was fabricated in the LioniX foundry using standard Si$_3$N$_4$ LPCVD while we deposited the TeO$_2$ layer using reactive radio frequency magnetron sputtering at ambient temperature, which is compatible with back end of line CMOS processing. More details on the fabrication process are given in [37]. Figures 1(a-c) show a cross-section, SEM image, and transverse electric-field (TE) fundamental mode profile of the designed waveguide calculated using a finite element method (FEM) mode solver, respectively. The hybrid waveguide core consists of a 1.6 μm × 0.4 μm Si$_3$N$_4$ strip coated with a 0.424-μm-thick TeO$_2$ film. The waveguide propagation loss is 0.6 dB/cm, measured by Q fitting 500-μm-diameter ring resonators on the same chip and with the same waveguide dimensions at a resonance wavelength of around 1550 nm [30]. The loss for the uncoated Si$_3$N$_4$ waveguide was measured to be 0.2 dB/cm using a similar approach, and the TeO$_2$ film loss was measured using prism coupling and found to be 2.45 ± 0.3 dB/cm and 0.48 ± 0.3 dB/cm for the fundamental TE mode at 638 nm and 1550 nm respectively. We suspect that the TeO$_2$ film loss might contribute to the excess loss in the hybrid waveguide and its propagation loss might be reduced by improving deposition conditions [30].

We calculated the waveguide dispersion using an FEM mode solver accounting for both the material and geometrical contributions. There are three materials in our waveguides, including SiO$_2$ lower cladding, Si$_3$N$_4$ core, and the TeO$_2$ coating. For SiO$_2$ and Si$_3$N$_4$ we used the Sellmeier equations previously used in [36]. For the TeO$_2$, we measured the refractive indices of the film using ellipsometry and fitted it to a Sellmeier model to obtain $n^2 - 1 = 2.8939\lambda^2/(\lambda^2 - 0.13967^2)$ for wavelengths from 400 nm to 1750 nm. Figure 1(d) shows the calculated dispersion profile of the fundamental TE mode for the uncoated and 424-nm TeO$_2$ coated 400-nm Si$_3$N$_4$ strip waveguide. It can be seen, as also shown in [36] for different waveguide dimensions, that by adding the TeO$_2$ coating layer to the normal dispersion thin Si$_3$N$_4$, anomalous dispersion can be achieved for our hybrid waveguide structure.

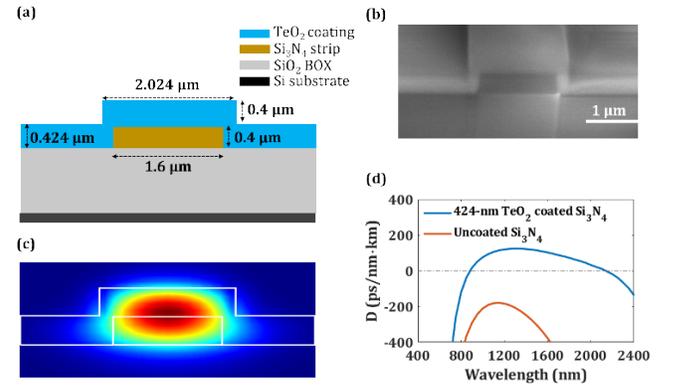

**Figure 1.** (a) Cross-section, (b) SEM image, and (c) TE mode profile of the TeO$_2$-coated Si$_3$N$_4$ waveguide, and (d) calculated waveguide dispersion profiles showing anomalous dispersion in the TeO$_2$-Si$_3$N$_4$ waveguide.

For SCG experiments, a 6.7-cm-long paperclip shaped waveguide was pumped by a Menlo fiber laser centered at 1565 nm with 100-fs pulses, 200-MHz repetition rate, and 76-mW average power. Figure 2(a) shows the experimental setup consisting of the femtosecond fiber laser free-space coupled to the waveguide through mirrors (M1, M2), a half-wave plate (H), a quarter-wave plate (Q), and a focusing lens (L). The lens-chip coupling loss was found to be 8 dB by measuring the insertion loss. We determined the coupling loss by first characterizing the loss per facet using identical lensed fibers (~3dB/facet) and then replacing one of the lensed fibers with the free-space focusing lens and recording the increase in insertion loss. The output was collected by a butt-coupled multimode fluoride fiber which was fed to two optical spectrum analyzers (OSAs). The first OSA (Ando) spans from 350 nm to 1750 nm and the second (Yokogawa) spans from 1200 nm to 2400 nm. Figure 2(b) shows an image of the chip and strong third harmonic generation (THG) across the visible wavelengths. Figure 2(c) shows the experimental results of the generated SC (red) and the input pump spectrum taken at low power without the chip in the setup (black). The maximum coupled average power was 10.94 mW, corresponding to a peak power of 482 W, for which the generated spectrum spans from 0.89 μm to 2.11 μm @ –20 dB level. Because the waveguide is in the anomalous dispersion regime, the SCG process is dominated by soliton fission and dispersive wave formation [6,7].

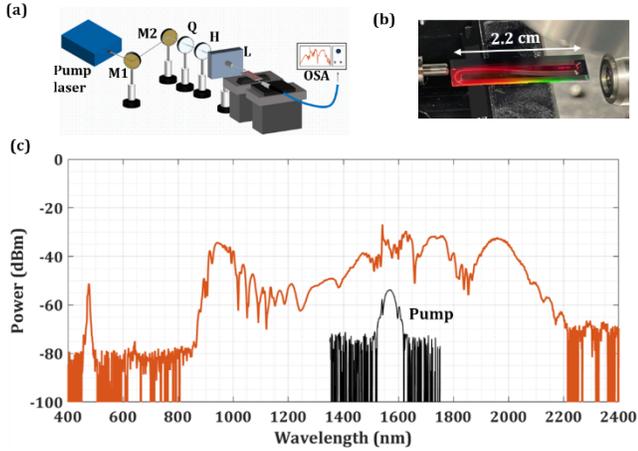

**Figure 2.** (a) Experimental setup, (b) generated SC (red), pump (black), and an image of the chip showing THG.

We varied the pump power and the corresponding generated SCs versus coupled peak power are plotted in Fig 3(a). The coupled power was varied using a broadband optical attenuator from 48.5. to 482 W. For an input power of 258 W, an octave-spanning SC is observed. This power corresponds to an average power of 5.86 mW and energy of 29.3 pJ.

Furthermore, we carried out a numerical simulation of the supercontinuum based on the generalized nonlinear Schrödinger equation [2]. The simulation results are shown in Figs. 3(b) and (c). In Fig. 3(b) we plot the output SC spectra for varying power like the experimental results in Fig. 3(a). To obtain this figure we initially used a calculated nonlinear parameter ($\gamma$) of 2.28 $W^{-1}m^{-1}$ using equations in [36]. We then varied this value from 1 to 3 $W^{-1}m^{-1}$ and found the simulated SCG best matches the experiment for a $\gamma$ value of 2.5 $W^{-1}m^{-1}$. We estimate an error of ±0.5 $W^{-1}m^{-1}$ contributed to by the uncertainties in the input pump coupling efficiency and pulse properties and variations in the SC propagation loss and output coupling efficiency with wavelength. We used this $\gamma$ value to calculate the $n_2$ of the $TeO_2$ and obtained a value of 1.4 × $10^{-18}$ $m^2/W$. This value agrees with that obtained in [38] and approximately 6 times higher than that of stoichiometric $Si_3N_4$ [21].

In Fig. 3(c) we show the spectral evolution of the SC along the waveguide length for the highest input power of 482 W. It is shown that soliton fission starts to take place at a relatively shorter length of only 0.6 cm. Hence, a significantly shorter waveguide could have been used to achieve similar spectral results. For the device under test, we measured an average output power coupled into the fiber of about 1 mW across the spectrum for a maximum average input power of 10.94 mW giving a 9.1% conversion efficiency. Since we assume lossless output coupling with the multimode fiber, this is a lower estimate of the conversion efficiency. Compared to the waveguide length used here, we estimate loss reductions of ~ 3.7 dB and several dB at the pump wavelength and short wavelength edge of the spectrum [37], respectively, for a waveguide length of 0.6 cm. Therefore, pumping a shorter $TeO_2$-$Si_3N_4$ waveguide can lead to an improved conversion efficiency with a significantly higher power across the spectrum to a level that the SCG can be considered useful as a broadband white light source for chip-scale applications. Furthermore, by adjusting the $TeO_2$ coating thickness to design the dispersion in the normal regime [36], highly broadband coherent supercontinuum generation [17] can also be explored on this platform.

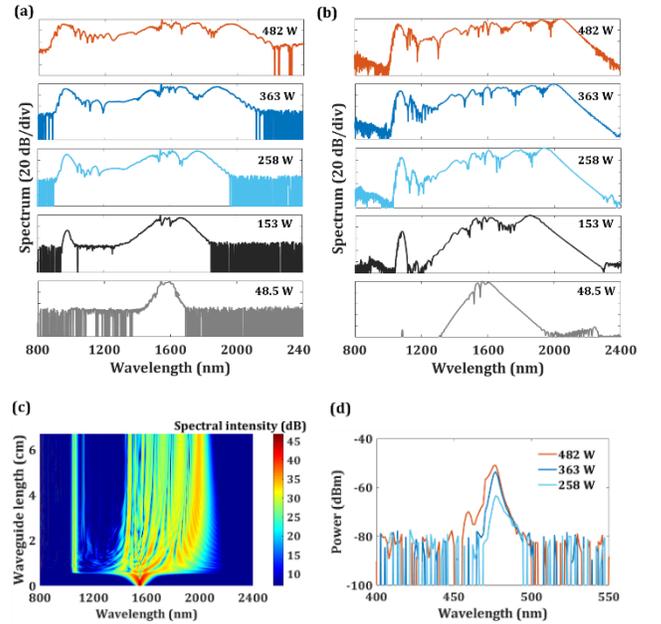

**Figure 3.** (a) The experimental and (b) simulated SC spectra at different pump powers, (c) spectrogram, and (d) zoomed-in third harmonic spectra for different powers.

In Fig. 3(d) we show a zoomed-in spectrum of the THG in our waveguides, which is similar to what we reported previously [33]. The FWHM level covers 12 nm bandwidth from 469 nm to 481 nm. Also, the peak of the spectrum is at the ~ −20 dB level of the entire SC. The average power is estimated to be 232 nW from the 10.94 mW input power. We notice that the measured spectrum only covers the blue bandwidth (450 nm to 500 nm), while colors including green, yellow, and red (~500 to 700 nm) are visible in the chip image in Fig. 2(b). This is likely due to the relatively higher propagation losses at shorter wavelengths, evident from the measured $TeO_2$ film loss of 2.45 ± 0.3 dB/cm for the fundamental TE mode at 638 nm. Hence most of the generated third harmonic signal is lost before reaching the output facet. Like SCG, the THG efficiency can be improved by using a shorter waveguide. This demonstration shows the potential of the hybrid platform for visible applications such as scanning displays and quantum information processing [39].

Table 1 shows a summary of broadband SCG demonstrated in various CMOS platforms with spectral coverage including NIR wavelengths. Our results compare well with most of the previous demonstrations in terms of propagation loss ($\alpha$) and pump pulse duration ($\tau_p$) and wavelength ($\lambda_0$). Our $TeO_2$-$Si_3N_4$ platform performs well in terms of peak power ($P_p$) requirement, with a useful octave-spanning SC bandwidth ($B_\lambda$) generated at a peak power of 258 W, the lowest compared to all other results obtained in the typical thick stoichiometric $Si_3N_4$ waveguides. Furthermore, we note that our waveguide length ($L$) is relatively longer than most of the other results in Table 1. However, we have shown in Fig. 3(c) that the same bandwidth and potentially higher output power can be achieved with a waveguide of > 0.6 cm. Lastly, among a few others, we also report strong THG.

**Table 1. Broadband NIR SCG in silicon photonic platforms**

| Platform | L (cm) | α (dB/cm) | $\tau_p$ (fs) | $\lambda_0$ (nm) | $P_p$ (kW) | $B_\lambda$, level (μm, dB) | Ref. |
|---|---|---|---|---|---|---|---|
| Silica | 350 | 0.00037 | 180 | 1330 | ~10.6 | 0.94–1.89, –5 | [4] |
| SOI | 0.5 | 1.5 | >50 | 1900 | 0.159 | 1.1–2.4, –20 | [7] |
| $Ta_2O_5$ | 0.5 | 0.1 | 80 | 1560 | ~9.9 | 0.7–2.4, –20 | [8] |
| USRN | 0.3 | 3 | 500 | 1555 | 0.34 | 1.06–2.24, –40 | [9] |
| SiN:D[*] | 0.5 | 0.5 | 74 | 1560 | 5.7 | 0.4–2.5, –45 | [17] |
| $Si_3N_4$ | 4.3 | 0.8 | 200 | 1300 | 0.704 | 0.67–2.03, –30 | [10] |
| $Si_3N_4$ | 0.8 | 0.7 | 92 | 1000 | ~0.346 | 0.67–1.94, –40 | [11] |
| $Si_3N_4$ | 0.6 | 0.5 | 120 | 1560 | 11.67 | 0.5–2.6, –30 | [12] |
| $Si_3N_4$[*] | 1 | - | 80 | 1550 | 11.7 | 0.5–3 | [13] |
| $Si_3N_4$[*] | 2 | 0.6 | 200 | 1510 | ~0.31 | 0.8–1.7 | [14] |
| $Si_3N_4$ | 0.5 | < 6 | < 90 | 1550 | ~6.7 | 0.4–4.5 | [15] |
| $Si_3N_4$ | 20 | 0.0209 | 50 | 1560 | 6.56 | 1.01–2.02, –20 | [16] |
| $Si_3N_4$ | 5.5 | - | 78 | 2100 | 3.84 | 1.2–2.9 | [18] |
| $Si_3N_4$[1] | 0.5 | - | 100 | 3200 | 90 | 0.5–4.44 | [19] |
| $Si_3N_4$-$TeO_2$[2] | 6.7 | 0.6 | 100 | 1565 | 0.258 | 0.94–1.93, –30 | This work |
|  |  |  |  |  | 0.482 | 0.89–2.11, –20 |  |

[*]Results where THG is also reported.

In summary, we have demonstrated octave-spanning SCG in a $Si_3N_4$ waveguide coated with a highly nonlinear $TeO_2$ film using a CMOS-compatible process. We show that an octave SC covering the entire telecom band spanning from 0.94 to 1.93 μm can be obtained at a low peak power of 258 W. For the maximum input power of 482 W, we measured an average on-chip output power of 1 mW. By fitting numerical calculations to the measured spectra, we estimate the nonlinear refractive index of the $TeO_2$ film to be 6 times that of stoichiometric LPCVD $Si_3N_4$. In addition, we observe THG over a wide range of visible wavelengths with a single pump source. The 400-nm-thick $Si_3N_4$ used here is readily incorporated in volume wafer-scale foundry processes and well-suited for linear photonic applications, while the $TeO_2$ can be doped with rare earth ions for active devices. Hence, the $TeO_2$-$Si_3N_4$ platform offers an alternate route to nonlinear photonics using thinner $Si_3N_4$ fabricated using conventional methods as well as a path to monolithic integration of linear, nonlinear, and active functionalities on silicon photonic chips. This work shows that $TeO_2$ can be used to both enhance the nonlinearity and provide a straightforward route to dispersion engineering in thin $Si_3N_4$, leading to efficient chip-scale nonlinear processes.

**Funding.** Ontario Ministry of Research and Innovation (ER17-13-077); Natural Sciences and Engineering Research Council of Canada (RGPIN-2017-06423); Canada Foundation for Innovation (35548); Mitacs Globalink Research Award (FR79012); Optica Amplify Scholarship; EU Horizon 2020 Framework Programme - Grant Agreement No.: 965124 (FEMTOCHIP), and Deutsche Forschungsgemeinschaft (SP2111) contract number PACE: Ka908/10-1.

**Acknowledgments.** The authors would like to acknowledge the Centre for Emerging Device Technologies (CEDT) at McMaster University for support with the sputtering system, LioniX International for assistance with layout and fabrication of the silicon nitride chips, and CMC Microsystems for the provision of products and services that facilitated this research, including Synopsis' RSoft Component Design Suite.

**Disclosures.** "The authors declare no conflicts of interest."

## REFERENCES

1. C. Lafforgue, M. Montesinos-Ballester, T.-T.-D. Dinh, X. Le Roux, E. Cassan, D. Marris-Morini, C. Alonso-Ramos, and L. Vivien, Photonics Res. **10**, A43 (2022).
2. J. M. Dudley, G. Genty, and S. Coen, Rev. Mod. Phys. **78**, 1135 (2006).
3. D. J. Jones, S. A. Diddams, J. K. Ranka, A. Stentz, R. S. Windeler, J. L. Hall, and S. T. Cundiff, Science **288**, 635 (2000).
4. D. Y. Oh, D. Sell, H. Lee, K. Y. Yang, S. A. Diddams, and K. J. Vahala, Opt. Lett. **39**, 1046 (2014).
5. L. Zhang, A. M. Agarwal, L. C. Kimerling, and J. Michel, Nanophotonics **3**, 247 (2014).
6. L. Yin, Q. Lin, and G. P. Agrawal, Opt. Lett. **32**, 391 (2007).
7. N. Singh, M. Xin, D. Vermeulen, K. Shtyrkova, N. Li, P. T. Callahan, E. S. Magden, A. Ruocco, N. Fahrenkopf, C. Baiocco, B. P.-P. Kuo, S. Radic, E. Ippen, F. X. Kärtner, and M. R. Watts, Light Sci. Appl. **7**, 17131 (2018).
8. K. F. Lamee, D. R. Carlson, D. R. Carlson, Z. L. Newman, Z. L. Newman, S.-P. Yu, and S. B. Papp, Opt. Lett. **45**, 4192 (2020).
9. Y. Cao, B.-U. Sohn, H. Gao, P. Xing, G. F. R. Chen, D. K. T. Ng, and D. T. H. Tan, Sci. Rep. **12**, 9487 (2022).
10. R. Halir, Y. Okawachi, J. S. Levy, M. A. Foster, M. Lipson, and A. L. Gaeta, Opt. Lett. **37**, 1685 (2012).
11. A. R. Johnson, A. S. Mayer, A. Klenner, K. Luke, E. S. Lamb, M. R. E. Lamont, C. Joshi, Y. Okawachi, F. W. Wise, M. Lipson, U. Keller, and A. L. Gaeta, Opt. Lett. **40**, 5117 (2015).
12. M. A. G. Porcel, F. Schepers, J. P. Epping, T. Hellwig, M. Hoekman, R. G. Heideman, P. J. M. van der Slot, C. J. Lee, R. Schmidt, R. Bratschitsch, C. Fallnich, and K.-J. Boller, Opt. Express **25**, 1542 (2017).
13. D. R. Carlson, D. D. Hickstein, A. Lind, S. Droste, D. Westly, N. Nader, I. Coddington, N. R. Newbury, K. Srinivasan, S. A. Diddams, and S. B. Papp, Opt. Lett. **42**, 2314 (2017).
14. Y. Okawachi, M. Yu, J. Cardenas, X. Ji, A. Klenner, M. Lipson, and A. L. Gaeta, Opt. Lett. **43**, 4627 (2018).
15. H. Guo, C. Herkommer, A. Billat, D. Grassani, C. Zhang, M. H. P. Pfeiffer, W. Weng, C.-S. Brès, and T. J. Kippenberg, Nat. Photonics **12**, 330 (2018).
16. A. Ishizawa, A. Ishizawa, K. Kawashima, R. Kou, X. Xu, T. Tsuchizawa, T. Aihara, K. Yoshida, T. Nishikawa, K. Hitachi, G. Cong, N. Yamamoto, K. Yamada, and K. Oguri, Opt. Express **30**, 5265 (2022).
17. I. Rebolledo-Salgado, Z. Ye, S. Christensen, F. Lei, K. Twayana, J. Schröder, M. Zelan, and V. Torres-Company, Opt. Express **30**, 8641 (2022).
18. E. Tagkoudi, C. G. Amiot, G. Genty, and C.-S. Brès, Opt. Express **29**, 21348 (2021).
19. Y. Fang, C. Bao, Z. Wang, B. Liu, L. Zhang, X. Han, Y. He, H. Huang, Y. Ren, Z. Pan, and Y. Yue, J. Lightwave Technol. **38**, 3431 (2020).
20. H. Zia, K. Ye, Y. Klaver, D. Marpaung, and K.-J. Boller, Laser Photonics Rev. **4**, 2200296 (2023).
21. K. Ikeda, R. E. Saperstein, N. Alic, and Y. Fainman, Opt. Express **16**, 12987 (2008).
22. M. H. P. Pfeiffer, A. Kordts, V. Brasch, M. Zervas, M. Geiselmann, J. D. Jost, and T. J. Kippenberg, Optica **3**, 20 (2016).
23. K. Wu and A. W. Poon, Opt. Express **28**, 17708 (2020).
24. J. S. Levy, A. Gondarenko, M. A. Foster, A. C. Turner-Foster, A. L. Gaeta, and M. Lipson, Nat. Photonics **4**, 37 (2010).
25. R. M. Grootes, M. Dijkstra, Y. Klaver, D. Marpaung, and H. L. Offerhaus, Opt. Express **30**, 16725 (2022).
26. V. A. G. Rivera and D. Manzani, eds., *Technological Advances in Tellurite Glasses* (Springer, 2017).
27. S. J. Madden and K. T. Vu, Opt. Express **17**, 17645 (2009).
28. K. Vu, S. Farahani, and S. Madden, Opt. Express **23**, 747 (2015).
29. K. Vu and S. Madden, Opt. Express **18**, 19192 (2010).
30. H. C. Frankis, H. M. Mbonde, D. B. Bonneville, C. Zhang, R. Mateman, A. Leinse, and J. D. B. Bradley, Photonics Res. **8**, 127 (2020).
31. K. Miarabbas Kiani, H. C. Frankis, H. M. Mbonde, R. Mateman, A. Leinse, A. P. Knights, and J. D. B. Bradley, Opt. Lett. **44**, 5788 (2019).
32. K. Miarabbas Kiani, H. C. Frankis, C. M. Naraine, D. B. Bonneville, A. P. Knights, and J. D. B. Bradley, Laser Photonics Rev. **16**, 2100348 (2022).
33. N. Singh, H. M. Mbonde, H. C. Frankis, E. Ippen, J. D. B. Bradley, and F. X. Kärtner, Photonics Res. **8**, 1904 (2020).
34. P. Muñoz, G. Micó, L. A. Bru, D. Pastor, D. Pérez, J. D. Doménech, J. Fernández, R. Baños, B. Gargallo, R. Alemany, A. M. Sánchez, J. M. Cirera, R. Mas, and C. Domínguez, Sensors **17**, 2088 (2017).
35. W. D. Sacher, J. C. Mikkelsen, Y. Huang, J. C. C. Mak, Z. Yong, X. Luo, Y. Li, P. Dumais, J. Jiang, D. Goodwill, E. Bernier, P. G.-Q. Lo, and J. K. S. Poon, Proc. IEEE. **106**, 2232 (2018).
36. H. M. Mbonde, H. C. Frankis, and J. D. B. Bradley, IEEE Photonics J. **12**, (2020).
37. H. C. Frankis, K. Miarabbas Kiani, D. B. Bonneville, C. Zhang, S. Norris, R. Mateman, A. Leinse, N. D. Bassim, A. P. Knights, and J. D. B. Bradley, Opt. Express **27**, 12529 (2019).
38. S.-H. Kim, T. Yoko, and S. Sakka, J. Am. Ceram. Soc. **76**, 2486 (1993).
39. Y. Lin, Z. Yong, X. Luo, S. S. Azadeh, J. C. Mikkelsen, A. Sharma, H. Chen, J. C. C. Mak, P. G.-Q. Lo, W. D. Sacher, and J. K. S. Poon, Nat. Commun. **13**, 6362 (2022).